\documentclass[12pt]{article} 
\pdfoutput=1

\usepackage[hyperfootnotes=false]{hyperref}
     \usepackage{amsmath}
      \usepackage{amssymb}
  \usepackage{graphicx}
  \usepackage{xcolor}
  \usepackage[utf8]{inputenc}
  \newlength{\abstractwidth}
 \usepackage{mciteplus} 

  \newcommand{\be}{\begin{equation}}
  \newcommand{\bea}{\begin{eqnarray}}
  \newcommand{\eea}{\end{eqnarray}}
  \newcommand{\ee}{\end{equation}}
  
  \newcommand{\es}[2] {\begin{equation} \label{#1} \begin{split} #2 \end{split} \end{equation}}
  \def\le{\left}
\def\ri{\right}
\newcommand\ov{\over}
\newcommand\sL{{\ensuremath{{\mathcal L}}}}
\newcommand\lam{\lambda}
\newcommand\Lam{\Lambda}
\newcommand\sig{\sigma}
\newcommand\ep{\epsilon}
\newcommand\al{\alpha}
\newcommand\p{\ensuremath{\partial}}
\newcommand{\vv}{v_{LC}}

 \numberwithin{equation}{section}
 \usepackage[normalem]{ulem}

\begin{document}

\begin{titlepage}
 % \rightline{}
  \bigskip

  \bigskip\bigskip

  \bigskip

\begin{center}
%\centerline
{\Large \bf 
On entanglement spreading in chaotic systems
}
 \bigskip
%\centerline
{\Large \bf { }} 
    \bigskip
\bigskip
\end{center}

  \begin{center}

 \bf {M\'ark Mezei$^1$ and Douglas Stanford$^2$ }
  \bigskip \rm
\bigskip
 
   $^1$Princeton Center for Theoretical Science, Princeton University, Princeton, NJ 08544 \\
\rm
 \bigskip
 $^2$Institute for Advanced Study,  Princeton, NJ 08540, USA

  \bigskip \rm
\bigskip
 
\rm

\bigskip
\bigskip

  \end{center}

 \bigskip\bigskip
  \begin{abstract}
\noindent We discuss the time dependence of subsystem entropies in interacting quantum systems. As a model for the time dependence, we suggest that the entropy is as large as possible given two constraints: one follows from the existence of an emergent light cone, and the other is a conjecture associated to the ``entanglement velocity'' $v_E$. We compare this model to new holographic and spin chain computations, and to an operator growth picture. Finally, we introduce a second way of computing the emergent light cone speed in holographic theories that provides a boundary dynamics explanation for a special case of entanglement wedge subregion duality in AdS/CFT.

 \medskip
  \noindent
  \end{abstract}
\bigskip \bigskip \bigskip

  \end{titlepage}

  %  \starttext \baselineskip=17.63pt \setcounter{footnote}{0}
   \tableofcontents

 % \sc

\section{Introduction}

In free field theory, the spreading of entanglement can be understood in terms of free-streaming particles \cite{Calabrese:2005in,Calabrese:2007rg,Cotler:2016acd}. In chaotic systems, this picture is no longer accurate \cite{Leichenauer:2015xra,Asplund:2015eha,Casini:2015zua}. We would like to propose a replacement, based on the idea that the entropy is as large as possible given two constraints: first, information must remain within an emergent light cone, and second, the rate of change of the entropy is bounded by a certain multiple of the area.

The analysis in this paper will revolve around three speeds that can be assigned to extended quantum systems. We introduce these now:
\begin{itemize}
\item[$v_E$] the ``entanglement velocity'' was defined in \cite{Hartman:2013qma,Liu:2013iza,Liu:2013qca} by the statement that at early times after a quench, the entropy of a large region grows as $\frac{d}{dt} S[A(t)] = v_E\, s_\text{th}\, \text{area}(A)$, where $s_\text{th}$ is the equilibrium entropy density.
\pagebreak
\item[$v_{LC}$] is the speed of the effective light cone, defined by some fixed and small threshold for commutators $\langle [W(t,x),V(0)]^2\rangle_\beta$, where $V,W$ are arbitrary operators that do not change the energy density drastically, and the expectation value is taken in the canonical ensemble with inverse temperature $\beta$.\footnote{The system does not need to be relativistic for this to make sense \cite{Lieb:1972wy}. Indeed, even in relativistic systems we often have $v_{LC}<c$, provided that we restrict to thermal-scale smeared operators that do not access the very high energy states where the true light cone is relevant.}
\item[$v_B$] the ``butterfly effect velocity'' was defined in \cite{Shenker:2013pqa,Roberts:2014isa} as the speed at which the region where the commutator is order one expands outwards. 
\end{itemize}
One expects a scrambling time delay between the effective light cone (where the commutator is small) and the ``butterfly cone'' (where it is order one). However, we suspect that generically the slopes are the same, so that $v_{LC}= v_B$. This is true in holography and likely to be true in a weakly coupled field theory, using an analysis along the lines of \cite{Stanford:2015owe}. We caution the reader that in the spin chain we study below it does not appear to be precisely true.  After the first version of this paper appeared,\footnote{Sentence added in v2.} \cite{Bohrdt:2016vhv} found that in the Bose-Hubbard chain $v_{LC}$ is significantly greater than $v_B$.   Although we have been careful to distinguish them in this introduction, we will use $v_B$ and $v_{LC}$ somewhat interchangeably in what follows. It would be interesting to refine our analysis in a way that distinguishes between the two velocities.

The main point of this paper will be to associate each of the speeds $v_E$ and $v_{LC}$ to a bound on the time dependence of the entropy, and to compare to data. The two bounds interact in a somewhat nontrivial way, and we believe that many coarse-grained features of entropy and information dynamics in systems with $v_{LC} = v_B$ can be captured by a simple model in which we simply saturate the combined bound. We will compare this model to new data from holography and from chaotic spin chain numerics. Our work should be understood as a combination of the tsunami picture \cite{Hartman:2013qma,Liu:2013iza,Liu:2013qca,Leichenauer:2015xra} and the models of \cite{Casini:2015zua,Ho:2015woa}.

The plan of the paper is as follows. In section \ref{boundsSec} we motivate the two bounds, one of which is essentially \cite{Hartman:2015apr}. In section \ref{holo} we show with new computations (presented in detail in \cite{Mezei:2016zxg}) that the actual holographic results are close to the combined upper bound. In particular, certain holographic systems saturate entropy as fast as allowed by the bounds, at least for spherical and strip entangling regions. In that section we also discuss bounds and holographic data for a mutual information quantity that tracks the spread of information in a more detailed way than the standard entropy. In section \ref{spin}, we make a similar comparison to numerical data from a simple chaotic spin chain.

In section \ref{growth} we discuss an operator growth model that saturates the entropy inequalities. In this model, $v_B=v_{LC}$ is the rate at which operators tend to grow, and $v_E$ is related to the small probability that they actually don't grow at all.

In section \ref{EEwedge}, we make a comment that relates the speed $v_B$ to a special case of entanglement wedge reconstruction. Finally, we conclude and provide a brief outlook. Further numerical data about the chaotic spin chain, the holographic computation relating $v_B$ to entanglement wedge reconstruction in higher derivative gravity theories, and the comparison of the bounds to the quasiparticle model are collected in the appendices.

\section{Two constraints on the time dependence of entropy} \label{boundsSec}

In this section we will argue for two inequality constraints on the time evolution of entropy. We assume that the state of interest has an approximately uniform energy density that allows us to assign an effective temperature $T = \beta^{-1}$. We will consider large entangling regions with characteristic size $r\gg\beta$, see \cite{Kundu:2016cgh} for a recent study of small regions.

\begin{figure}[!h]
\begin{center}
\includegraphics[scale=.55]{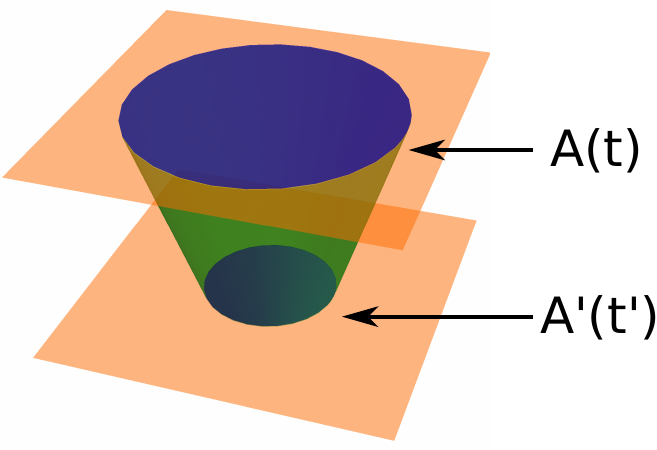} \hspace{1cm}
\includegraphics[scale=.6]{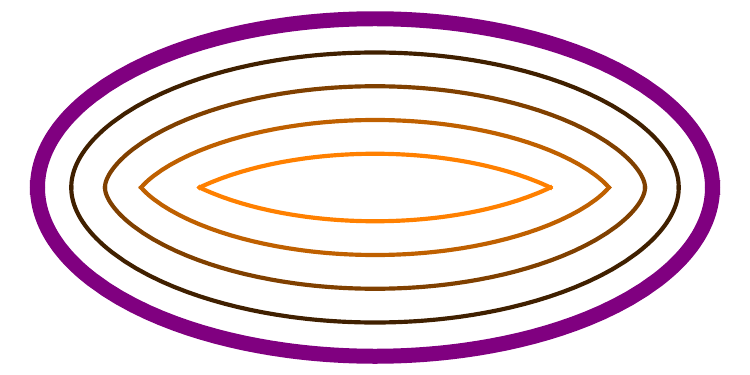}
\caption{{\bf Left:} The  ``$\vv$ cone'' for a disk $A'(t')$ passes through $A(t)$.  Any information in $A'(t')$ is contained in $A(t)$, hence $A'(t')$ is a subsystem of $A(t)$ and we can use the monotonicity of relative entropy \eqref{relent}.
{\bf Right:} The tsunami wavefront for different times corresponding to the case where $\p A$ is an ellipse (purple). Lighter color means later time. 
}\label{butterflyLC}
\end{center}
\end{figure}

The first constraint comes from the idea that the system will have an emergent light cone at speed $\vv$. Concretely, this means that operators will approximately commute with other operators if their spatial separation is more than $\vv$ times their time separation. This imposes a speed limit for entanglement \cite{Casini:2015zua,Hartman:2015apr} by an argument exactly parallel to that of \cite{Hartman:2015apr}. The argument goes as follows: any subsystem $A'(t')$ at time $t'<t$ is actually a subsystem of $A(t)$ at a later time $t$ provided that the future ``$\vv$ cone'' of $A'(t')$ passes entirely through $A(t)$, see figure \ref{butterflyLC}.

Monotonicity of the relative entropy then implies
\be\label{relent}
S(\rho_A(t)|\rho_A^{(\text{th})}) \ge S(\rho_{A'}(t')|\rho_{A'}^{(\text{th})})\,,
\ee
where $\rho_A^{(\text{th})}$ is the reduction of the thermal density matrix (at the effective temperature of our state) to region $A$.
It was pointed out by \cite{Hartman:2015apr} that if the region $A$ is much larger than the thermal scale $\beta$, then $S(\rho_A|\rho_A^{(\text{th})})\approx S[\rho_A^{(\text{th})}] - S[\rho_A]$, so \eqref{relent} implies
\be\label{con3b}
I_1[A(t)] \ge I_1[A'(t')]\,, \hspace{20pt} I_1[A(t)] \equiv S[\rho_A^{(\text{th})}] - S[\rho_A(t)]\,. 
\ee
We can understand this inequality intuitively. $I_1[A(t)]$ is a measure of the amount of information contained in the density matrix $\rho_A(t)$. The inequality is just the statement that if the subsystem $A'(t')$ is contained in $A(t)$, then it contains no more information than $A(t)$ does. The parameter that enters here is $\vv$, the slope of the effective light cone. This will depend on the system (and, in general, state) of interest. However, it can be established by an independent calculation, such as the commutator of local operators. If we consider this bound alone, we get 
\es{TsunamiVolume}{
S[A(t)]-S[A(0)]\leq s_{\text{th}}\cdot \text{vol}(\text{tsunami}(t))\,,
}
where $S[A(0)]$ subtracts the entropy of the initial state to make the LHS finite even in field theory, and $\text{vol}(\text{tsunami}(t))$  is the volume covered by a tsunami wave  in time $t$ propagating in from $\p A$ with speed $\vv$ \cite{Hartman:2015apr}, see figure \ref{butterflyLC}.\footnote{We have also assumed $S[\rho_A^{(\text{th})}]\approx s_{\text{th}}\cdot \text{vol}(A) + S[A(0)]$ in deriving \eqref{TsunamiVolume}.}

The second inequality we propose is that the time derivative of the entropy of any region $A$ should be bounded by the area:
\be\label{con1}
\frac{d}{dt}S[A(t)] \leq v_E\, s_{\text{th}}\cdot \text{area}(A)\,,
\ee
where $s_{\text{th}}$ is the equilibrium thermal entropy density at the energy/temperature scale we are working.\footnote{Here we consider states that are approximately energy-thermalized, meaning that the local energy density is approximately uniform. Otherwise one could take a state that locally had very high energy. In this situation one expects the above inequality to be violated, at least if $s_{\text{th}}$ refers to the global system.} The parameter $v_E$ is defined by requiring equality in \eqref{con1} for large regions and short times after a quench.\footnote{This is expected to be well-defined in the sense that $v_E$ does not depend on the region. The region independence can be explicitly seen in holography \cite{Hartman:2013qma,Liu:2013iza,Liu:2013qca} and in the quasiparticle model describing free field theories \cite{Casini:2015zua,Cotler:2016acd}.}  So the content of \eqref{con1} is that the entropy should never change faster than right after a quench. Unlike \eqref{con3b}, this is simply a conjecture. There are rigorous bounds of the form \eqref{con1} but with a coefficient that depends on the Hilbert space dimension and operator norms \cite{2006PhRvL..97e0401B,2014arXiv1411.0680M}. We expect that in practice there might be a better bound that applies even to theories with unbounded operators, provided that the energy is not too high. \eqref{con1} is a guess for what this bound might be. Moreover, \eqref{con1} can be proven in holographic theories \cite{Mezei:2016zxg}.

We will now make three comments about the above constraints. First, as noted by \cite{Hartman:2015apr}, in the setup of a global quench, \eqref{con3b} implies \eqref{con1} with $v_E$ replaced by the light cone speed. (In \cite{Hartman:2015apr} this was taken to be $c$.) It follows that
\be\label{velessvb}
v_E\leq \vv\,.
\ee
Indeed, for the particular example of charge neutral high-temperature states of holographic theories with boundary spacetime dimension $d$, we have \cite{Hartman:2013qma,Liu:2013iza,Liu:2013qca,Shenker:2013pqa,Roberts:2014isa}
\be\label{holspeeds}
v_E = \frac{\sqrt{d}(d-2)^{\frac{1}{2} - \frac{1}{d}}}{[2(d-1)]^{1 - \frac{1}{d}}}\,, \hspace{20pt}v_{LC} = v_B = \sqrt{\frac{d}{2(d-1)}}\,, \hspace{20pt} d \equiv \substack{\text{field theory} \\ \text{spacetime dim}}\,,
\ee
which satisfy \eqref{velessvb}. In \cite{Mezei:2016zxg} the null energy condition is used to verify \eqref{velessvb} for general states in holographic theories. In the spin chain we will discuss in section \ref{spin}, we have $v_B \approx v_{LC} \approx 2 v_E$. 

Second, if we know the entropies of all subregions at a given time, we can use \eqref{con3b} and \eqref{con1} together to give upper bounds on the entropy at later times (or equivalently, lower bounds on the information). We will see that in many cases, the optimal bound involves both conditions. It is interesting to compare this bound with the actual answer for the entropy. In the next section we will make this comparison for holographic theories and for a simple chaotic spin chain.

Third, the bounds inform us about the saturation time $t_S$, when the entropy reaches the thermal value. For general shapes, it is somewhat nontrivial to evaluate the best bound that follows from \eqref{con3b} and \eqref{con1}. But it is easy to see that we will have the inequalities
\be\label{satBound}
t_S\geq {r_\text{insc}\ov \vv }\,, \hspace{20pt} t_S\geq {\text{vol}(A)\ov v_E\, \text{area}(A)}\,.
\ee
Here $r_{\text{insc}}$ is the radius of the largest ball that fits inside $A$. The first of \eqref{satBound} is a better bound for ``round'' shapes, and is easy to derive from \eqref{TsunamiVolume}. The second is better for elongated shapes and follows from integrating \eqref{con1}.

\section{Comparing the bounds to data}\label{holo}
\subsection{Global quench}
We begin by studying the entropy $S[A(t)]$ following a global quench from a sparsely entangled state. Then we set the initial conditions $S[A(0)] = 0$ for all regions $A$, and we compute the upper bound on $S$ that follows from \eqref{con3b} and \eqref{con1}.\footnote{Here we are implicitly subtracting the UV-sensitive area contribution. To be more precise we should study $I_1[A(t)]$ instead, which does not contain the area term. However, we will continue with $S[A(t)]$ to make better contact with previous work.} An interesting test case is to consider $A$ to be a sphere of radius $r_A$. Then we use \eqref{con3b} with $A'$ chosen to be a sphere of radius $0\leq r\leq r_A$ at time $t' = t - (r_A-r)/\vv>0$. This gives
\be
S[A(t)] \leq  S[A'(t')] +  s_{\text{th}}\cdot(\text{vol}(A) - \text{vol}(A'))\,, \hspace{20pt} t' = t - (r_A-r)/\vv\,.
\ee
We further bound $S[A'(t')]$ using \eqref{con1}. All together,
\be\label{tomin}
S[A(t)] \leq v_E\, s_{\text{th}} \cdot \text{area}(A')\,t' + s_{\text{th}}\cdot(\text{vol}(A) - \text{vol}(A'))\,.
\ee
To get the best upper bound, we minimize \eqref{tomin} over $0\leq r\leq r_A$, finding
\es{boundS}{
S[A(t)] &\leq v_Es_{\text{th}}\,\text{area}(A)\,t \hspace{20pt} t \leq t_0\equiv r_A\frac{\vv -v_E}{(d-2)\vv v_E}\\
S[A(t)]&\leq s_{\text{th}}\,\text{vol}(A)\left[1 - \frac{v_E}{\vv }\left(\frac{(d-2)v_E}{(d-1)v_E - \vv }\right)^{d-2}\left(1 - \frac{\vv t}{r_A}\right)^{d-1}\right] \hspace{20pt} t \ge t_0
}
before saturation.
\begin{figure}[!h]
\begin{center}
\vspace{10pt}
\includegraphics[scale=.41]{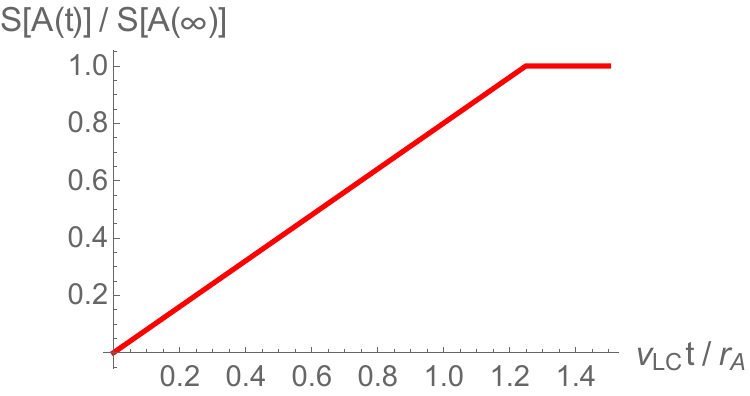}
\includegraphics[scale=.41]{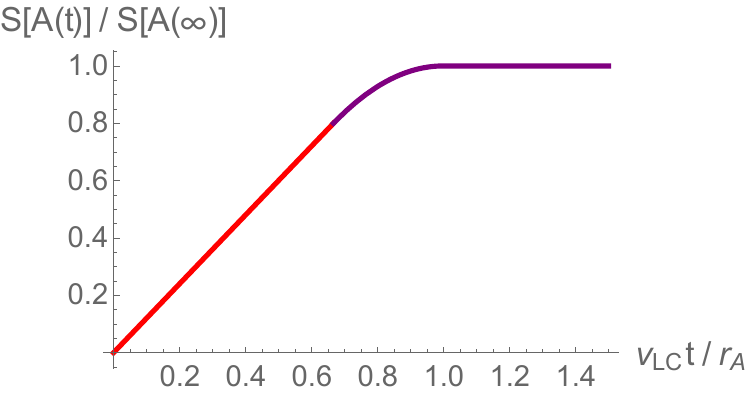}
\includegraphics[scale=.41]{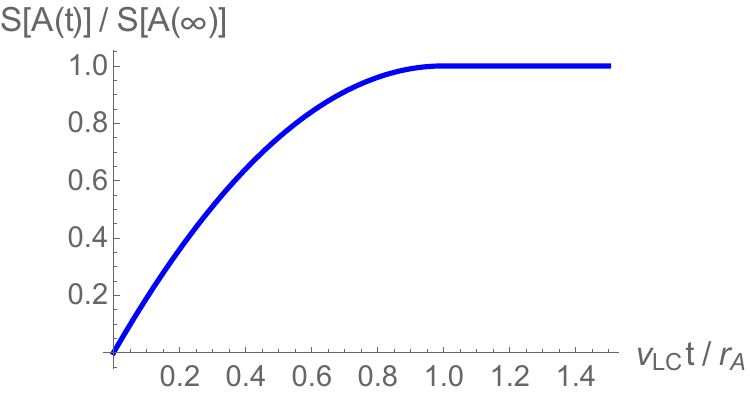}
\caption{The bound \eqref{boundS} for $v_E<\frac{\vv}{d-1}$ (left) where only the $v_E$ constraint plays a role, $ \frac{\vv}{d-1}<v_E<\vv$ (middle) where both are important, and $v_E = \vv$ (right) where only the $\vv$ constraint is relevant. In the middle plot the change of color indicates the position of $t_0$ defined in \eqref{boundS}.}\label{3regimes}
\end{center}
\end{figure}

\begin{figure}[!h]
\begin{center}
\includegraphics[scale=.55]{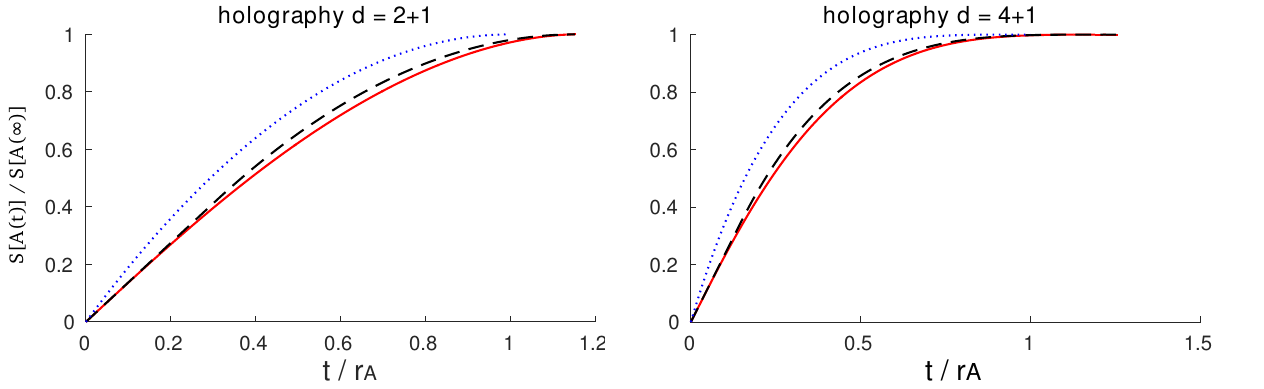}
\caption{We plot $S[A(t)]$ until saturation for a large sphere. Black/dashed is our upper bound using the speeds \eqref{holspeeds}, blue/dotted is the bound from relativistic causality \cite{Hartman:2015apr} and red/solid is the holographic computation \cite{Mezei:2016zxg}. The dashed lines should be understood as one-parameter ($v_E$) fits, because $\vv$ is determined by an independent calculation based on commutators (or equivalently, out-of-time order correlation functions). }\label{Hcomparison}
\end{center}
\end{figure}

If $v_E < \frac{\vv}{d-1}$, then the $\vv$ constraint plays no role and we have linear growth all the way until saturation, see figure \ref{3regimes}. However, if $\frac{\vv}{d-1}<v_E<\vv$, then the bound depends in an important way on both constraints. This is the case for holographic theories with $d>2$. In figure \ref{Hcomparison}, we plot the bound \eqref{boundS} together with new holographic computations \cite{Mezei:2016zxg} of $S[A(t)]$ for a large ($r_A\gg \beta$) ball-shaped region. The actual holographic answer lies just a little below the upper bound, and saturation happens as fast as allowed by the first inequality in \eqref{satBound}.\footnote{There is a small subtlety in the $2+1$-dimensional case. It is shown in \cite{Mezei:2016zxg} that in this case saturation happens slightly later than the soonest time the bound would allow, and the time derivative of the entropy at saturation is non-zero. The effect is invisible on  figure \ref{Hcomparison}, and is related to the issue of ``discontinuous" saturation in holography, discussed briefly in section \ref{EEwedge} and in detail in \cite{Mezei:2016zxg}. \label{foot7}}

It is also straightforward to consider shapes other than spheres following the logic of \eqref{tomin}. For strips only the bound  \eqref{con1} is relevant before the saturation time, and we get
\es{StripBound}{
S[A(t)] &\leq v_E\,s_{\text{th}}\cdot\text{area}(A)\,t\,.
}
This bound is saturated for strips in holographic theories for all times, so the saturation time bound \eqref{satBound} applicable for elongated regions is also saturated \cite{Hartman:2013qma,Liu:2013iza,Liu:2013qca}. It would be interesting to work out the upper bound for other shapes.

\subsection{The thermofield double state}
The increase of entropy after a quench is a result of the delocalization of information under time evolution. We can map this out in more detail by considering another entropy quantity, the two-sided mutual information \cite{Morrison:2012iz,Hartman:2013qma} in the thermofield double state
\be
|TFD(t)\rangle = Z(\beta)^{-1/2}\sum_{n}e^{-\beta E_n/2 - iE_n t}\,|n\rangle_L|n\rangle_R\,.
\ee
The state $|TFD(0)\rangle$ is a thermal version of a maximally entangled state, with a high degree of entanglement between large subsystems $A_R\subset R$ and the corresponding subsystem $A_L\subset L$. 

It will be convenient to think about $R$ as the physical system, evolving in time, and $L$ as a static reference system that keeps track of the movement of information in $R$. In other words, we view the time evolution as $|TFD(t)\rangle = e^{-iH_R t}\,|TFD(0)\rangle$, where $H_R$ is the Hamiltonian acting on the $R$ system. The key point is that the mutual information
\be\label{mutdef}
I[B_R(t),A_L] \equiv S[B_R(t)] + S[A_L] - S[B_R(t)\cup A_L]
\ee 
can be understood as a measure of the amount of information originally contained in $A_R(0)$ that is now contained in $B_R(t)$. Note that this is a statement about only the physical $R$ part of the system; the thermofield double is a tool to make this notion precise.

We will think about $I[B_R(t),A_L]$ for the case where $A,B$ are concentric balls, of radius $r_A$ and $r_B$. The mutual information is then a function of three parameters: $t,r_A,r_B$, which we take to be large compared to the inverse temperature $\beta$. It is interesting to fix $t,r_A$ and consider $I$ as a function of $r_B$. This tells us how much of the information originally contained in a ball of radius $r_A$ is later contained in a concentric ball of radius $r_B$.

\begin{figure}[!h]
\begin{center}
\includegraphics[scale=.55]{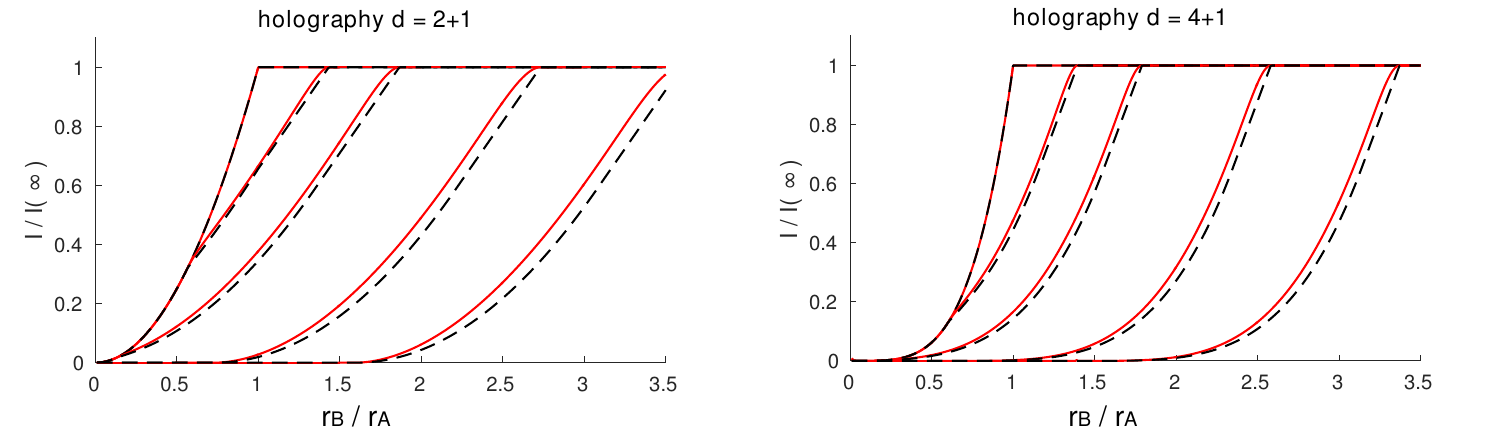}
\caption{We plot $I[B_R(t),A_L]$ as a function of $r_B/r_A$. The different curves correspond to $t/r_A = 0,0.5,1,2,3$. Dashed/black is the lower bound, solid/red is the holographic result.}\label{HInfcomparison}
\end{center}
\end{figure}

In the $|TFD(t)\rangle$ state, the first two terms in \eqref{mutdef} are exactly given by the thermal values. The third term can be computed in holography by applying the Ryu-Takayanagi (RT) formula \cite{Ryu:2006bv} and the techniques of \cite{Mezei:2016zxg}. We can also compute a bound on this quantity using an analog of \eqref{con3b} for mutual information, namely
\be\label{emlc}
I[B_R(t_1),A_L(t_2)]\ge I[B_R'(t_1'),A_L'(t_2')]
\ee
for all $A'(t_2'),B'(t_1')$ causally contained (in the $\vv$ sense) in $A(t_2),B(t_1)$. This inequality, together with \eqref{con1}, and the initial conditions $I[A_L(0),B_R] = 2s_{\text{th}}\cdot \text{vol}(A\cap B)$ lead to the lower bound on $I$,
\be\label{tomaximize}
I[A_L(t),B_R]\ge 2s_{\text{th}}r^{d-1}\frac{\Omega_{d-2}}{d-1} - v_Es_{\text{th}}r^{d-2}\Omega_{d-2}\left(t - \frac{r_B+r_A-2r}{\vv}\right)
\ee
for any $0\leq r \leq \text{min}\{r_A,r_B\}$ such that the expression in parentheses is positive.  We get the best bound by maximizing over such $r$ values. In the above expression $\Omega_{d-2}$ is the volume of the unit $S^{d-2}$. The maximization can be done explicitly, but rather than giving the answer in the various cases, we will just plot the result against the holographic answer. We show this in figure \ref{HInfcomparison}. The holographic answer always lies a little above the lower bound, but the curves are quite close. This means that the holographic answer is almost as small as it can be, given $\vv$ and $v_E$.

\begin{figure}[!h]
\begin{center}
\includegraphics[scale=.8]{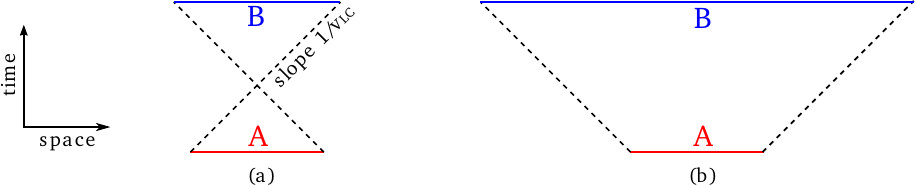}
\caption{If $B$ is smaller than shown in (a), the lower bound and the holographic answer indicate that $B$ contains none of the information in $A$. If $B$ is larger than shown in (b), it contains all of the information in $A$. The dashed lines are $\vv$ cones, not light cones.}\label{infFig}
\end{center}
\end{figure}

As a function of $r_B$, there are two interesting points: the value where the mutual information first becomes nonzero, and the value where it saturates. In both the bound and in holography, these are given in terms of $\vv$ as
\be\label{2eqs}
r_B^{\text{first nonzero}} = \vv\, t - r_A\,, \hspace{20pt} r_B^{\text{saturates}} = \vv\, t + r_A\,.
\ee
This is illustrated in figure \ref{infFig}. (Here, we are assuming $v_E \geq \frac{\vv}{d-1}$.) It is clear that the mutual information should saturate for $r_B \ge \vv \,t + r_A$, since then the emergent light cone implies that $B(t)$ contains $A(0)$ as a subsystem, see figure \ref{infFig}. However, the fact that the mutual information is zero until $r_B = \vv \,t - r_A$ is not obvious, and is a striking difference from a free field theory.\footnote{For this to hold, we have assumed that the holographic dual black hole geometry is such that in the quench setup giving the same final state black hole the entropy saturates ``continuously". This is not true for the $2+1$-dimensional case, but the violation of the first equality in \eqref{2eqs} is invisibly small on figure \ref{infFig}. (The second equation in \eqref{2eqs} remains valid.)  More discussion of these issues is found in footnote \ref{foot7}, section \ref{EEwedge}, and  \cite{Mezei:2016zxg}, where examples with large violations of $r_B^{\text{first nonzero}} = \vv\, t - r_A$ and $t_S=r/v_B$ are also discussed. }

\subsection{Spin chain data}\label{spin}
\begin{figure}[!h]
\begin{center}
\includegraphics[scale=.52]{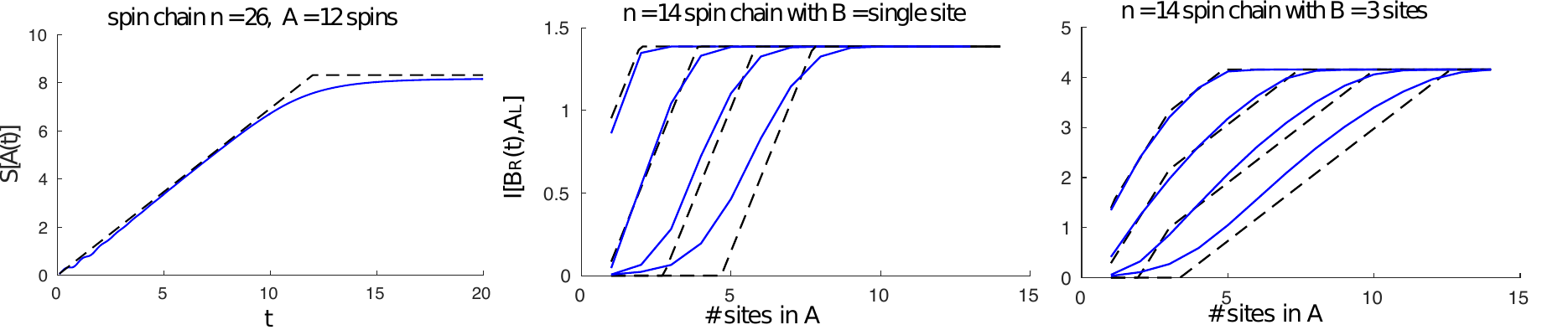}
\caption{{\bf Left:} we plot $S[A(t)]$ for the spin chain with initial state $|Y+\rangle$ and $n = 26$ spins. The region $A$ consists of the first 12 spins. We find $v_E = 1.0$. {\bf Middle, right:} we plot $I[A(t),B_R]$ for the spin chain along with the lower bounds with $\vv = 1.85$ (see appendix \ref{renyiapp}) and $v_E = 1.2$ (which is the best fit for the maximally entangled state). The different curves correspond to different times $t$, increasing from left to right. }\label{SCInfcomparison}
\end{center}
\end{figure}

In this section we repeat the analysis of the previous sections for a simple chaotic spin chain with the Hamiltonian
\be\label{spinH}
H = \sum_{i}\Big\{\sigma_z^{(i)}\sigma_z^{(i+1)} - 1.05\,\sigma_x^{(i)} + 0.5\, \sigma_z^{(i)}\Big\}\,.
\ee
First we consider the evolution of the entropy following a ``quench.'' This was considered for essentially the same model by \cite{Kim:2013bc}. By representing the Hamiltonian as a sparse matrix and evolving the state directly, we can avoid exact diagonalization and study a reasonably long chain of $n = 26$ spins. We take the initial state $|Y+\rangle$, which thermalizes well \cite{Banuls:2011str}. The entropy shows clear linear growth until near saturation, with a fitted slope of $v_E \approx 1$ in lattice units, see figure \ref{SCInfcomparison}. We note that if we had studied a smaller subsytem, the saturation would have been somewhat sharper. R\'enyi entropies are plotted in appendix \ref{renyiapp}.

We can also consider the mutual information quantity \eqref{mutdef}. In this $d = 2$ setting, the maximization of \eqref{tomaximize} over $r$ is simple, and we find the lower bound
\be
I[A(t),B_R] \ge \text{max}\le\{0,\,2s_{\text{th}}\,\text{min}\{L_A,L_B\} - v_Es_{\text{th}}\, \text{max}\le\{0,\,t - \frac{|L_A-L_B|}{\vv}\ri\}\ri\}\,,
\ee
where $L_A,L_B$ are the respective lengths of the $A,B$ subsystems, which are both assumed to begin at the end of the spin chain. Notice that this bound involves both $v_E$ and $\vv$ in a nontrivial way. To compute the mutual information numerically, we use exact diagonalization to study an infinite temperature thermofield double of 14 spins on each side. Note that this is just the standard maximally entangled state. See figure \ref{SCInfcomparison} for a comparison to the lower bound from \eqref{tomaximize}. Although the agreement is not as good as for holography, the actual mutual information is surprisingly close to the lower bound. It seems that the spin chain result starts to lag behind somewhat as we move to later times. This may be related to the fact that $v_{LC}$ appears numerically to be slightly larger than $v_B$ for this system.

\section{An operator growth model}\label{growth}

In this section we discuss an operator growth model that saturates inequalities analogous to \eqref{con3b}, \eqref{con1} for R\'enyi  2-entropy. The model is not derived from unitary time evolution, but it may help to give an understanding of $v_E$ and $v_B=v_{LC}$. The discussion is inspired by the model of \cite{Ho:2015woa}, but the details are different.\footnote{The operator counting model of \cite{Ho:2015woa} gives $v_E=\vv$, and in $d>2$ it gives a saturation time that violates \eqref{satBound}; our model improves on these aspects.}

The setting is as follows. We imagine that we have a system of $N$ qubits arranged in a regular $(d-1)$ dimensional lattice. They start in the ``quench'' pure state $|0\rangle|0\rangle...|0\rangle$ so that the density matrix of the complete system is
\be
\rho(0) = \prod_i \frac{1 + \sigma_z^{(i)}}{2} = 2^{-N}\sum_{\substack{\text{subsets } S \\ \text{of sites}}} \  \prod_{i \in S}\sigma_{z}^{(i)}\,.\label{term}
\ee
We would like to compute $\text{tr}_A\,\rho_A(t)^2$. We can think about computing this by studying the time evolution of an individual term in \eqref{term}
\be
\prod_{i\in S}\sigma_z^{(i)}(t) = \sum_{\substack{\text{subsets }\tilde{S}\text{ of sites} \\ \text{and Pauli indices}}} \  c_{S\rightarrow \tilde{S}}(t)\prod_{(i,\alpha) \in \tilde{S}} \sigma^{(i)}_\alpha\,.\label{sum}
\ee
Here we are expanding the evolution of a given term in $\rho$ as a sum of products of Pauli operators. The R\'enyi 2-entropy can be obtained from 
\es{pur1}{
\text{tr}_A\,\rho_A(t)^2 &=\frac{1}{2^{2N}}\text{tr}_A\left\{\le(\text{tr}_{A^c}\sum_{S_1}\prod_{i\in S_1}\sigma_z^{(i)}(t)\ri) \ \le( \text{tr}_{A^c}\sum_{S_2}\prod_{j\in S_2}\sigma_z^{(j)}(t)\ri)\right\}\\
&=\frac{1}{2^{2N}}\sum_{S_1,S_2}\sum_{\tilde{S}_1,\tilde{S}_2} c_{S_1\rightarrow \tilde{S}_1}(t) c_{S_2\rightarrow \tilde{S}_2}(t) \, \text{tr}_A\left\{\le(\text{tr}_{A^c}\prod_{(i,\al)\in \tilde{S}_1}\sigma_\al^{(i)}\ri) \ \le( \text{tr}_{A^c}\prod_{(j,\beta)\in \tilde{S}_2}\sigma_\beta^{(j)}\ri)\right\}\,,
}
where $A^c$ is the complement of $A$, and we plugged in the definitions \eqref{term}, \eqref{sum}. Because the Pauli operators are traceless,  if $\tilde{S}_{1,2}$ contains sites outside $A$, then tracing over $A^c$ gives zero. This restricts the sum to $\tilde{S}_{1,2}\subset A$, and we can use the orthogonality of Pauli operators to obtain:
\es{pur2}{
\text{tr}_A\,\rho_A(t)^2 &= \frac{1}{2^{|A|}}\sum_{S_1,S_2}\sum_{\tilde{S}\subset A}c_{S_1\rightarrow \tilde{S}}(t)c_{S_2\rightarrow \tilde{S}}(t) \approx \frac{1}{2^{|A|}}\sum_{S}\sum_{\tilde{S}\subset A}c_{S\rightarrow \tilde{S}}(t)^2\,,
}
where in the final step we neglected the off-diagonal contributions. For large regions, this can be justified under the assumption that the coefficients $c_{S\rightarrow \tilde{S}}$ have random and uncorrelated signs.\footnote{ In detail, let ${\cal S}$ and $\tilde{\cal S}$ denote the sizes of the sets from which we take $S_{1,2}$ and $\tilde{S}$ respectively. Then there are ${\cal S} \, \tilde{\cal S}$ diagonal, and ${\cal S}^2 \, \tilde{\cal S}$ off-diagonal terms.  If we assume independent fluctuating signs for the off-diagonal terms, their contribution is proportional to $\pm {\cal S} \, \sqrt{\tilde{\cal S}}$, which can be neglected compared to the diagonal terms:
\es{Dominance}{
{\cal S} \, \sqrt{\tilde{\cal S}}\ll{\cal S} \, \tilde{\cal S} \,.
 } }
 Fixing $S$ in the final sum corresponds to focusing on a single product operator in the expansion \eqref{term}; the contribution of such a term can be understood as the ``probability'' that this operator remains inside the region $A$ after time $t$. Here, in defining the probability we are thinking of the space of operators as a Hilbert space, with basis vectors the different Pauli strings. The Heisenberg evolution of an operator defines a quantum evolution on this state space.

A simple model of this the probability is to assume that operators grow  at a rate $v_B = v_{LC}$: the probability is $1$, if the support of the operator (growing outwards from the initial support with $v_B$  for time $t$ ) is inside $A$, and $0$, if the support reached outside. Hence all operators supported inside the ``dry region" not reached by the tsunami discussed around \eqref{TsunamiVolume} contribute to the sum \eqref{pur2} with weight $1$, and the rest of the operators have $0$ weight. Then  \eqref{pur2} becomes 
\es{pur3}{
\text{tr}_A\,\rho_A(t)^2 &\approx \exp\le[-s_{\text{th}}\, \text{vol}(\text{tsunami}(t))\ri]\,
}
where $s_{\text{th}}$ is $\log 2$ times the density of sites. This model leads to an entropy saturating the inequality \eqref{TsunamiVolume} for all times.\footnote{We note that the tensor network inspired toy model of \cite{Casini:2015zua} also leads to the saturation of this inequality.}

Next, we introduce a refinement of the simple operator growth model discussed above. We assume that operators mostly grow outwards, at a rate $v_B = v_{LC}$. However, we also assume that there is some tail in the probability distribution for a given operator to remain the same size. This tail decays exponentially in time with a coefficient proportional to the area of the boundary of the operator: $e^{-\gamma\, t\cdot\text{area}}$. As an example, we can compute the probability that an operator initially filling a ball of radius $r<r_A$ will be within a concentric ball of radius $r_A$ after time $t$. The optimal strategy is for the operator to pay to stay the same size for a certain time $t'$ and then grow outward until time $t$, which leads to
\be
P(r,r_A,t) = e^{-\gamma\,  \text{max}(0,t')\cdot\text{area}(r)}, \hspace{20pt} t' \equiv t - \frac{r_A - r}{v_B}\,.
\ee
We now use this formula to evaluate \eqref{pur2} for the case that $A$ is a ball of radius $r_A$. We imagine that we are evaluating the sum for large regions, so we use a continuum notation. The concentric ball-shaped operators dominate this sum \eqref{pur2}, and we find
\be
\text{tr}\,\rho_A(t)^2 \approx e^{-s_{\text{th}}\text{vol}(r_A)}\left[1 + \int_0^{r_A} dr\ P(r,r_A,t) \, e^{s_{\text{th}}\text{vol}(r)}\right]\,,
\ee
where the one is a special treatment of the identity operator, which does not grow. The R\'enyi 2-entropy is $S_2[A(t)] = -\log \text{tr}\,\rho_A(t)^2$. For large regions the integral is dominated by the maximum, and we recover the maximization over $r$ of \eqref{tomin}. In other words, this model saturates the entropy bounds for spherical regions. More specifically, we find that it saturates the bounds with $v_E = \frac{\gamma}{s_{\text{th}}}$.

Thus, the operator growth model provides a picture of entanglement spread that saturates the bounds. In this model, $v_B = v_{LC}$ is the rate at which operators mostly grow, and $v_E$ is related to the small probability that operators actually do not grow. For spherical regions we get two complementary pictures of saturation: from the perspective of the bounds saturation occurs when the tsunami covers the whole entangling region, while in the operator picture saturation happens when the support of an initially small operator in the center grows bigger than the region. The agreement between the operator growth model and the bounds reinforces our confidence in the overall picture provided in this paper.

\section{The emergent light cone and the entanglement wedge} \label{EEwedge}
In this section, we will give a second discussion of the emergent light cone from the perspective of information spreading. Rather than studying the dynamics of information following a global quench, we will examine the delocalization of a small perturbation to the equilibrium thermal state. The calculation is very simple, but it connects in an interesting way to the somewhat mysterious entanglement wedge \cite{Czech:2012bh,Wall:2012uf,Headrick:2014cta} subregion duality \cite{Bousso:2012sj} in AdS/CFT.

Concretely, the problem is the following: imagine acting with a light local operator $W_x$ on the thermal state. Initially, some information about which operator we applied can be recovered from a local measurement at position $x$. But as time passes, scrambling delocalizes the information over a larger and larger region. We can think about this as the growth of the operator $W_x(-t)$, where its size is just the smallest region that contains significant information about the applied operator. We will define $\tilde{v}_B$ as the rate of growth as a function of time. One expects to find $\tilde{v}_B = v_B = v_{LC}$.

In a theory with a gravity dual, we can evaluate this size using subregion duality, which asserts that certain subregions of the boundary theory completely describe corresponding subregions of the bulk. The initial thermal state is represented by a static black hole, and acting with $W_x$ introduces a small perturbation that falls towards the horizon. The smallest subregion of the boundary that contains significant information about $W_x$ after time $t$ is simply the smallest boundary subregion such that the corresponding bulk subregion contains most of the falling particle's wave function. We will follow \cite{Czech:2012bh,Wall:2012uf,Headrick:2014cta} and assume that the bulk subregion is the ``entanglement wedge.'' For our purposes, this is just the region of a constant time slice of the bulk that is contained within the RT surface associated to the boundary subregion. 

\begin{figure}[!h]
\begin{center}
\includegraphics[scale=.65]{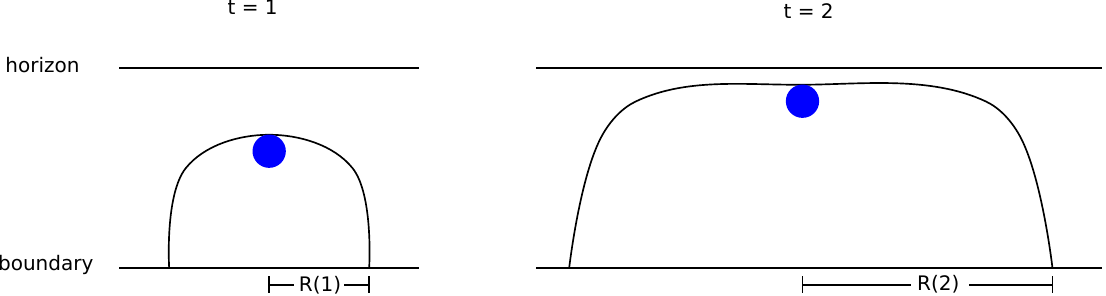}
\caption{We show two different time slices of the geometry outside the horizon of the black hole. The position of the falling particle (blue dot) is shown at each time, along with the (growing) minimal entanglement wedge that contains it. }\label{vbEEfig}
\end{center}
\end{figure}
Our analysis will focus on the near horizon region, where it is convenient to use a Rindler radial coordinate $\rho$ that measures proper distance from the horizon. The near-horizon metric of a general static planar black hole is then
\be\label{rindler}
ds^2 = -\left[\rho^2 + O(\rho^4)\right]\left(\frac{2\pi}{\beta}\right)^2dt^2 + d\rho^2 + \left[r_h^2 + 2\pi r_h \rho^2/\beta + O(\rho^4)\right]\frac{dx^idx^i}{\ell_{AdS}^2}\,,
\ee
where $\beta$ is the inverse temperature and $r_h$ is the area-radius of the horizon. In Schwarzschild or Rindler coordinates, the falling wave packet never crosses the horizon, but it is easy to check that as time advances it approaches exponentially,
\be\label{rad}
\rho(t) = \rho_0 \ e^{-\frac{2\pi}{\beta}t}\,.
\ee

To evaluate the size of $W_x(t)$, we need to find the smallest boundary region such that the RT surface extends down to the radius in \eqref{rad}. The optimal choice is to take a ball-shaped region of radius $R$, on a constant $t$ slice in the boundary. The RT surface can then be parameterized by $\rho(x^i)$, and by symmetry, it will depend on only one (radial) coordinate in the boundary. Finding the surface exactly would involve a nonlinear ODE, but for large boundary regions, most of the RT surface lies very close to the horizon, where the equation linearizes.\footnote{One can solve for the full surface analytically  in a double expansion \cite{Liu:2013una}, but the precision of \eqref{area} is enough for our purposes.} Indeed, near the horizon one can check that
\be\label{area}
\text{Area} = (r_h/\ell_{AdS})^{d-3}\int d^{d-1}x \left[(r_h/\ell_{AdS})^2 + \frac{1}{2}(\partial_i \rho)^2+\frac{1}{2}\mu^2\rho^2 + O(\rho^4)\right]\,,
\ee
where $\mu^2 = 2\pi r_h(d-1)/\beta\ell_{AdS}^2$.

Solutions to the equation of motion $\partial_i^2\rho = \mu^2\rho$ arising from \eqref{area} will vary exponentially as a function of $x^i$. Suppose that the radius of closest approach to the horizon is $\rho_{min}$, and that it happens at the origin of the $x^i$ coordinates. Then the solution in the near-horizon region is \cite{Liu:2013una} 
\be
\rho(x^i) = \rho_{min}\,\frac{\Gamma(a+1)}{2^{-a}\mu^{a}}\,\frac{I_{a}(\mu |x|)}{|x|^{a}} \hspace{20pt} a = (d-3)/2\,.
\ee
Eventually $\rho$ will exceed $\beta$ and the surface will exit the near-horizon region. From this point, it will reach the boundary within an order one distance in $x$. We can therefore determine the size of the operator $R$ in terms of $\rho_{min}$, to within an order one error, by solving
\be\label{min}
\beta = \rho_{min}\,\frac{\Gamma(a+1)}{2^{-a}\mu^{a}}\,\frac{I_{a}(\mu R)}{R^{a}} \hspace{20pt} \implies \hspace{20pt} \rho_{min} \approx e^{- \mu R}\,,
\ee
where we ignored prefactor powers and constants in the approximate expression.

Now we put these two steps \eqref{rad} and \eqref{min} together. In order for the falling particle to be within the entanglement wedge, we need $\rho_{min} \leq \rho(t)$, which implies that $R \ge \tilde{v}_B t$ with
\be\label{speed}
\tilde{v}_B = \frac{2\pi}{\beta \mu} = \sqrt{\frac{2\pi\ell_{AdS}^2}{(d-1)\beta\, r_h}}\,.
\ee
For the special case of an uncharged black hole in Einstein gravity, we get $\tilde{v}_B = \sqrt{d/2(d-1)}$. This matches \eqref{holspeeds}, so we see that indeed $\tilde{v}_B = v_B$. We will make four further comments:

{\it (1)} It can be checked that \eqref{speed} agrees with $v_B$ derived from the shock wave computations for a general black hole background in Einstein gravity. We are not certain if $\tilde{v}_B = v_B$ in general, but in appendix \ref{higherder} we show that $\tilde{v}_B = v_B$ remains true in certain higher derivative theories, where we know how to do both computations. Also, the fact that saturation of $I[B_R(t),A_L]$ saturates at the point predicted by $v_B$ in the left panel of figure \ref{SCInfcomparison} is an indication that $\tilde{v}_B = v_B$ in the spin chain as well.

 {\it (2)} Although we have presented this calculation as a second derivation of $v_B$, one can turn the logic around and view these results as a dynamical explanation for a special case of entanglement wedge subregion duality \cite{Czech:2012bh,Wall:2012uf,Headrick:2014cta}. Other proposals have been made for the bulk region that is described by a given boundary region, including the ``causal wedge'' \cite{Bousso:2012sj,Czech:2012bh,Hubeny:2012wa}. That proposal would have led to $\tilde{v}_B = 1$.

Let us clarify this point slightly. In order to get an operator deep in the bulk, we can start with a local operator $W_x$ and time evolve. To measure this operator in the CFT at a later time $t$, relativistic causality implies that we need no more than a ball of radius $t$ in the field theory. If this were the only constraint, we would end up with reconstruction only within the causal wedge. However, because of the emergent $v_B = v_{LC}$ cone, we can measure $W_x(-t)$ on a smaller subregion, of radius $v_B t$. We've seen that the specific value of $v_B$ singles out the entanglement wedge. Notice that the boundary domain of dependence of this smaller subregion will nowhere contain $W_x(-t)$ as a local operator, but it will approximately contain it as a nonlocal operator. This possibility was previously conjectured to be relevant for subregion duality, at the very end of \cite{Bousso:2012mh}.

{\it (3)} We can consider non-spherical regions of the boundary, and ask whether they can be used to reconstruct the operator. It is an easy extension of the above to check that the falling particle will be within the entanglement wedge if the region contains a ball of radius $v_B t$ surrounding the initial location of the operator. (More precisely, this is correct up to an imprecision of scale $\beta$ in the boundary coordinate.)\footnote{Corrected in v3: we thank Ying Zhao and Henry Lin for giving a counterexample to an ``if and only if'' claim in the original version.}

{\it (4)} Finally, we can also view the above calculation as giving the saturation time for the entropy of a ball-shaped region following a global quench in certain circumstances. The argument is simple. In the model of a quench discussed in \cite{Liu:2013iza,Liu:2013qca} an infalling shell of matter creates the black hole, and we get a nontrivial time dependence for the entropy because the Hubeny-Rangamani-Takayanagi (HRT)  \cite{Hubeny:2007xt,Dong:2016hjy} surfaces cross the shell. In the case of ``continuous" saturation, saturation of entropy occurs at time $t_S$, when the HRT surface climbs out from the AdS region behind the shell, and just barely touches the infalling matter. At this time, the geometry that the HRT surface is experiencing is that of a static black hole.

Note that at this point, the HRT surface becomes the RT surface that we have been discussing above. The role of the trajectory of the falling particle is played by the infalling null shell. It follows that the saturation time is $t_S = R / \tilde{v}_B$.\footnote{In \cite{Liu:2013iza,Liu:2013qca} a saturation velocity $c_E=R/t_S$ was introduced to characterize the saturation time; in this language $c_E=\tilde{v}_B$. We are grateful to Dan Roberts for pointing out the connection to $c_E$.} In the holographic cases we have been able to analyze, $\tilde{v}_B = v_B = v_{LC}$, so we saturate the first bound in (\ref{satBound}). It is interesting that the two setups appear superficially different: the falling particle is localized in the boundary theory spatial directions and its back reaction can be neglected, while the null shell is translation invariant and it creates the black hole geometry. For the problem we consider, however, the only thing that matters is that both follow the trajectory \eqref{rad}, and that the HRT surface only experiences the static black hole part of the geometry.

However, in certain black hole geometries it can happen that  there are multiple extremal surfaces corresponding to a given boundary theory time, and according to the HRT prescription we have to pick the one with minimal area. In such a case, the saturation is ``discontinuous": the HRT surface barely touching the null shell doesn't play a role in saturation, $t_S>R/v_B$, and the time derivative of the entropy at saturation is discontinuous. The conditions under which this happens is analyzed in detail in \cite{Mezei:2016zxg}. For uncharged black holes, saturation is discontinuous in $d=3$ and continuous in $d>3$.

\section{Discussion}
Our motivation for this work was to clarify the relationship between $v_E$, the ``entanglement velocity'' and $v_B$, the ``butterfly velocity.'' We have argued that they are associated to two different bounds on the time evolution of the entropy.

These are not universal bounds, in the sense that they depend on parameters $v_E$ and $v_{LC}$ that will vary from system to system. But we found evidence from holography and (to a somewhat lesser extent) spin chain numerics that the true evolution of the entropy is well approximated by saturating the combined bound. We suggest that for chaotic systems with $v_B = v_{LC}$, this may be an appropriate replacement for the particle streaming picture of entropy dynamics in free field theories \cite{Calabrese:2005in,Casini:2015zua,Cotler:2016acd}.

{\it Added in v2:} While we expect that generically $v_B = v_{LC}$, this does not appear to be precisely true  in the spin chain that we studied. In the recent exciting paper \cite{Bohrdt:2016vhv}, it was found that  $v_{LC}$ is significantly greater than $v_B$ in the Bose-Hubbard chain. We suspect that these violations are due to the special kinematics  in one spatial dimension, but the relation between the two velocities (or more generally the shape of the commutator $\langle [W(t,x),V(0)]^2\rangle_\beta$) deserves further investigations. When $ v_{LC}>v_B $, we have to use $v_{LC}$ in the bound  \eqref{con3b}. Because a significant amount of quantum information spreads with the slower speed $v_B$, the entropy is not expected to lie as close to the combined bounds, as in the case $v_B = v_{LC}$. 
 
It would be nice to have a more precise understanding of $v_E$. In the holographic dual, $v_E$ seems to be an interesting quantity, as it is determined by the geometry behind the horizon of a black hole. One hint comes from the operator growth model discussed in section \ref{growth}, in which $v_B$ is related to the typical speed at which operators grow as a function of time, and $v_E$ is related to the small probability that they do not grow.  Another possibility \cite{Hartman:2013qma,Hosur:2015ylk}, is that a tensor network perspective may be helpful.

It would also be nice to know whether there are subregion shapes for which the holographic answer for the entropy is far from saturating the combined bounds discussed in this paper.

\bigskip
\noindent{\bf Acknowledgements} 

\noindent We are grateful to J.~Camps, T.~Grover, D.~Huse, H.~Liu, X.~Qi, and D.~Roberts for discussions. The research of M.M. was supported in part by the U.S. Department of Energy under grant No. DE-SC0016244. D.S. is supported by the Simons Foundation grant 385600.

\appendix

\section{More spin chain numerics}\label{renyiapp}
\begin{figure}[!h]
\begin{center}
\includegraphics[scale = 0.65]{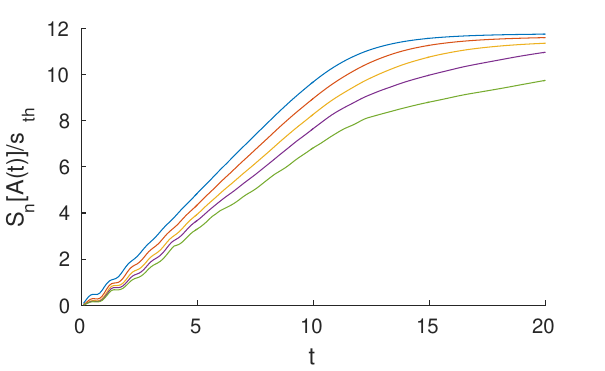}
\includegraphics[scale = 0.65]{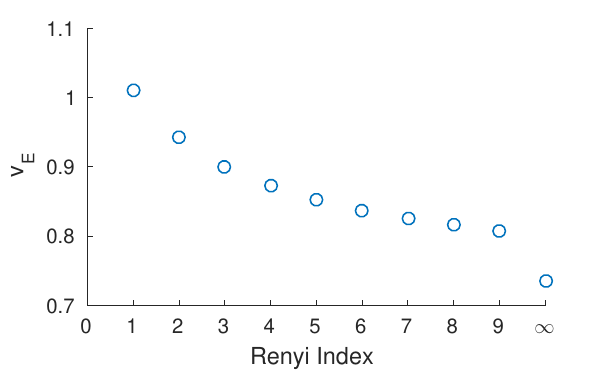}
\caption{At left we show the  entropy of the first 12 of 26 spins, starting in the product state $|Y+\rangle$. The different curves represent the von Neumann entropy (top), and the 2,4,8,$\infty$ R\'enyi entropies. At right we indicate fits to the initial slope. }\label{Renyis}
\end{center}
\end{figure}

In this appendix we present some further numerical results in the chaotic spin chain defined by the Hamiltonian \eqref{spinH}. First we discuss the growth of the R\'enyi entropies following a quench (see also \cite{GroverUnpublished}) from the initial product state $|Y+\rangle$. We evolve the state by directly applying the Hamiltonian to the state as a sparse matrix, avoiding the need for exact diagonalization. This allows us to study a 26 site chain, and we evaluate the entropy of the first 12 sites. The results are plotted in figure \ref{Renyis}. If we had studied a smaller subsytem, the saturation (particularly of the higher R\'enyis) would have been somewhat sharper.

We note that $v_E$ seems to depend on the R\'enyi index. This is reasonable given our interpetation (see section \ref{growth}) of $v_E$ as being determined by a small tail in the probability distribution for the evolution of an operator. Although we have not attempted the calculation, it seems likely that $v_E$ would depend on the R\'enyi index in holography as well, since the geometry will be affected by the backreaction of the R\'enyi brane.

\begin{figure}[!h]
\begin{center}
\includegraphics[scale = 0.7]{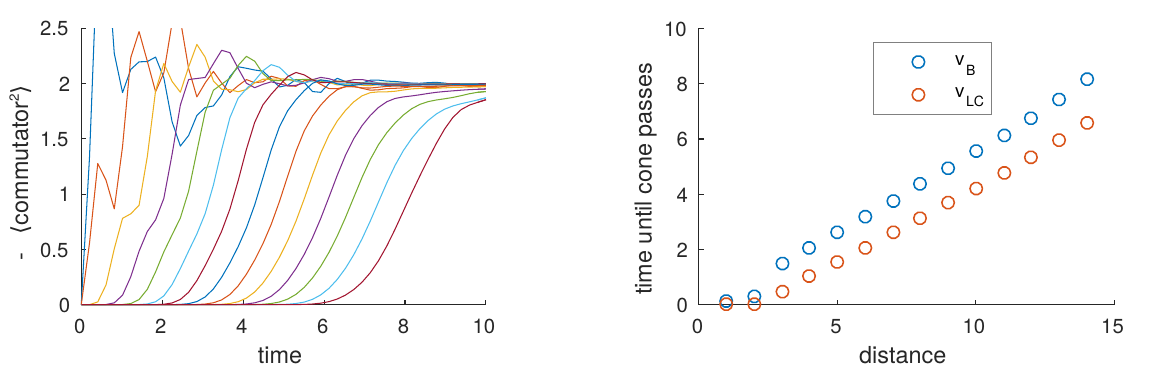}
\caption{At left we show the commutator $-\langle [\sigma_y^{(1)}(t),\sigma_y^{(k)}]^2\rangle$ for $k = 1...14$, in the infinite temperature state. At right we show the times at which the commutators for the different values of $k$ pass $1$ ($v_B$ cone) and $0.1$ ($v_{LC}$ cone). Linear fits between $3,13$ give $v_{LC} = 1.85$ and $v_B = 1.7$.}\label{spinChainSpeedsFig}
\end{center}
\end{figure} 
Next, we discuss the computation of $v_{LC}$ and $v_B$ in the spin chain. We choose to define $v_{LC}$ by a value of the commutator of 0.1, and we choose to define $v_B$ as a value of $1$. For these choices we find $v_B \approx 1.7$ and $v_{LC} \approx 1.85$. The results do seem to depend (weakly) on these specific choices. The results are plotted in figure \ref{spinChainSpeedsFig}.

\section{Higher derivative corrections} \label{higherder}

In this appendix we show that the method of computing $v_B$ from shock waves and $\tilde v_B$ from the entanglement wedge agrees in four derivative gravity theories to all orders in the higher curvature couplings. These are theories where we not only know the generalization of the RT formula, but also know the equation of motion that it satisfies \cite{Dong:2013qoa}. The Lagrangian for the theories we study is 
\es{4derRT}{
\sL&=R+\lam_1R^2+ \lam_2 R_{\mu\nu}R^{\mu\nu}+\lam_3 R_{\mu\nu\rho\sig}R^{\mu\nu\rho\sig}\\
&= R + \Lam_\text{GB}\le(R_{\mu\nu\rho\sig}R_{\mu\nu\rho\sig}-4 R_{\mu\nu}R^{\mu\nu} +R^2\ri)+\Lam_{1}\,R^2+\Lam_2 \, R_{\mu\nu}R^{\mu\nu}\,.
}
In the second line we have reexpressed the couplings in a way that simplifies some of the expressions below.

Before we discuss the calculation, we should mention a limitation of our analysis. Higher derivative theories require completion by massive states such as strings \cite{Camanho:2014apa}. Because the butterfly velocity comes from a high energy scattering problem, it may receive corrections due to Regge behavior of strings, in addition to the effective field theory higher derivative corrections. This did not happen at the order in $\alpha'$ studied in \cite{Shenker:2014cwa}, but it may happen at the next order. We have not attempted to check this, but we caution the reader that Regge corrections may disturb the agreement we find between $v_B$ and $\tilde{v}_B$ below.

\subsection{Higher derivative corrections to $v_B$ from shock waves}
The general static planar black hole metric can be written as
\es{Metric}{
ds^2= {1\ov z^2}\le[-{f(z)\ov h(z)} dt^2+ d\vec{x}^2+{dz^2\ov f(z)}\ri]\,,
}
where we put the horizon at $z=1$, hence $f(1)=0$. $h(z)$ is a positive function, and we allow arbitrary matter to support the geometry.

The shock wave profile can be determined in higher curvature gravity in a straightforward manner following the methods developed in \cite{Sfetsos:1994xa}, and recently concisely summarized in \cite{Roberts:2016wdl}. Let us write the equations of motion corresponding to \eqref{4derRT} as:
\es{4derEOM}{
G_{\mu\nu}+\lam_1 H^{(1)}_{\mu\nu}+\lam_2 H^{(2)}_{\mu\nu}+\lam_3 H^{(3)}_{\mu\nu}=8\pi G_N\, T_{\mu\nu}\,,
}
where we treat the negative cosmological constant as matter, and $H^{(i)}_{\mu\nu}$ can be found e.g. in \cite{Sinha:2010pm}. The black hole \eqref{Metric} solves this equation of motion with matter $\bar T_{\mu\nu}$. We now add a shockwave on top of this background. It will be convenient to work in Kruskal coordinates and look for the solution:
\es{Kruskal}{
\overline{ds}^2&=2 A(uv) \,dudv+ B(uv) dx^2\\
ds^2&=\overline{ds}^2-2 A(uv) h(x) \delta(u) du^2\\
T&=\bar T+ \delta T\\
\delta T&=\le[{\cal E} \delta(x)- 2 h(x)\bar T_{uv}\ri]\delta(u) du^2 - 2 h(x) \bar T_{vv} \delta(u) du dv\,,
}
where the contributions from $\bar T$ in $\delta T$ have to be included because we are using discontinuous coordinates \cite{Sfetsos:1994xa}. Plugging this Ansatz into \eqref{4derEOM} we get an equation for the shock wave profile $h(x)$. In terms of the couplings defined in \eqref{4derRT}, the equation of motion is
\es{hEq}{
&\le[\le(1+C_1(\Lam_1,\Lam_2)\ri)\p^2-{B'(0)\ov A(0) }\le({d-1\ov 2}+C_2(\Lam_1,\Lam_2)\ri)+ \Lam_2\,\p^4\ri]h(x)= {{\cal E}\ov A(0)} \delta(x)\,,
}
where the coefficients $C_i(\Lam_1,\Lam_2)$ depend on the black hole metric, and we used that $B(0)=1$.\footnote{Their explicit form is given by
\es{hEqExpl}{
&C_1(\Lam_1,\Lam_2)\equiv-4\le({A'(0)\ov A(0)^2}+{(d-1)B'(0)\ov A(0)}\ri)\, \Lam_1-{(d+1)B'(0)\ov A(0)}\, \Lam_2\\
&C_2(\Lam_1,\Lam_2)\equiv\le({2(d-1)A'(0)\ov A(0)^2}-{(d-1)(3d-10)B'(0)\ov A(0)}+{4\le[3A'(0)^2-2A(0)A''(0)-2(d-1)A(0)^2 B''(0)\ri]\ov A(0)^3 B'(0)}\ri)\, \Lam_1\\
&\qquad\qquad\qquad+\le(-{3(d-1)(d-3)B'(0)\ov 4A(0)}+{2\le[3A'(0)^2-2A(0)A''(0)-2(d-1)A(0)^2 B''(0)\ri]\ov A(0)^3 B'(0)}\ri)\, \Lam_2\,.
}}
Note that the Gauss-Bonnet coupling constant $\Lam_\text{GB}$ doesn't appear in the equation. However, $v_B$ still receives corrections in Gauss-Bonnet gravity, all of which are encoded in the change of the black hole solution.\footnote{E.g. in a Schwarzshild black hole $v_B$ is modified to \cite{Roberts:2014isa}
\be
v_B(\Lam_\text{GB})=N_\#\, \sqrt{d\ov 2(d-1)}\,, \hspace{20pt} N_\#^2\equiv \frac12 \le(1+\sqrt{1-4\Lam_\text{GB}}\ri)\,.
\ee
}

Using \eqref{speed}, the speed $v_B$ can be read off from the tradeoff between the exponential growth of scattering of the blueshifted particles $e^{\frac{2\pi}{\beta}t}$ and the decay of the shockwave profile $e^{-\mu|x|}$ at large $x$ \cite{Shenker:2013pqa,Roberts:2014isa}.
The decay constant $\mu$ can be determined by plugging into the fourth order equation  \eqref{hEq}. We get a second order equation for $\mu^2$:
\es{muSq}{
\le(1+C_1(\Lam_1,\Lam_2)\ri)\mu^2-{B'(0)\ov A(0) }\le({d-1\ov 2}+C_2(\Lam_1,\Lam_2)\ri)+ \Lam_2\,\mu^4=0\,.
}
 This equation has two solutions for $\mu^2$. We take the branch which gives back the Einstein result $\mu^2={(d-1)\ov 2}\,f_1$, when we take $\Lam_i\to0$. (The other branch diverges as we take $\Lam_2\to0$.)

For comparison with the entanglement wedge computation of $\tilde v_B$ it will be useful to convert from the Kruskal coordinates expression \eqref{hEqExpl} to the coordinates used in \eqref{Metric}:
\es{CCoeff}{
f(z)&\equiv f_1 (1-z)+f_2(1-z)^2+f_3(1-z)^3+\dots\,, \qquad h(z)\equiv h_0+h_1 (1-z)+h_2(1-z)^2+\dots\,,\\
{B'(0)\ov A(0) }&=f_1\,, \qquad 
C_1(\Lam_1,\Lam_2)=- \le[\le(4d-{3 h_1\ov h_0}\ri)\,f_1+4f_2\ri]\,\Lam_1-\le(d+1\ri)\,f_1\,\Lam_2\,,\\
C_2(\Lam_1,\Lam_2)&=-\le[\le(3d(d-1)-{5d-9\ov 2}\,{h_1\ov h_0}+{3h_1^2\ov h_0^2}-{5h_2\ov h_0}\ri)\,f_1+\le(6(d-1)-{3h_1\ov h_0}\ri)\,f_2+6f_3\ri]\,\Lam_1\\
&-\le[\le(\frac34(d+1)(d-1)-(d-2){h_1\ov h_0}+{3h_1^2\ov 2h_0^2}-{5h_2\ov 2h_0}\ri)\,f_1+\le(2(d-1)-{3h_1\ov 2h_0}\ri)\,f_2+3f_3\ri]\,\Lam_2\,.
}

\subsection{Higher derivative corrections to the entanglement wedge}

We now determine the entanglement wedge in higher derivative gravity. The static RT surface for a sphere is given by the function $z(r)=1-\ep \, s(r)^2$, where $\ep$ is small corresponding to close approach to the horion.\footnote{In section \ref{EEwedge} Rindler coordinates were used for the same problem, here we found it more convenient to work in $z$ coordinates. Introducing the variable $s(r)$ instead of its square simplifies our equations.} In \cite{Dong:2013qoa} a generalized area formula for proposed for higher derivative theories (see also \cite{Camps:2013zua}), but the equation of motion satisfied by the RT surface has up to now only been shown to follow from this area function in four derivative and Lovelock theories (see \cite{Camps:2014voa} for further discussion on the equation of motion). We will concentrate on four derivative gravity as in \eqref{4derRT}, where the generalized area is:
\es{4derRT2}{
S&={1\ov 4 G_N}\int dy\ \sqrt{g}\le[1+2\lam_1 R+\lam_2\le(R^a_{\,\,a}-\frac12 K^a K_a\ri)+\lam_3\le(R^{ab}_{\,\,\,\,\, ab}-K^{a\mu\nu}K_{a\mu\nu}\ri)\ri]\,.
} 
We construct the geometric quantities entering this functional below. The Euler-Lagrange equation from \eqref{4derRT2} determines the RT surface.

Let us take a spherically symmetric surface $z(r)$. Because it is a codimension-2 surface, it has two orthogonal normal vectors:
\es{NormalVec}{
n^{(1)}&={z\ov \sqrt{f(z)\ov h(z)}}\, \le(1,0,\dots \ri)\,, \quad
n^{(2)}={z\ov \sqrt{\le(z'\ri)^2+{1\ov f(z)}}}\,  \le(0,-z',0\dots,1 \ri)\,.
}
From these we can construct the quantities needed for \eqref{4derRT2}
\es{NormalVec2}{
P_{\mu\nu}&=g_{\mu\nu}-\eta_{ab}n^{(a)}_\mu n^{(b)}_\nu\,,\quad 
K_{a\mu\nu}=\frac12 \, P^{\al}_{\,\,\mu}\, P^{\beta}_{\,\,\nu}\,\sL_{n^{(a)}} g_{\al\beta}\,, \quad K_{a}=g^{\mu\nu}K_{a\mu\nu}\,,\\  R_{ab}&=n^{(a)\mu}n^{(b)\nu}n^{(c)\rho}n^{(d)\sig}R_{\mu\nu\rho\sig}\,,\quad R_{abcd}=n^{(a)\mu}n^{(b)\nu}R_{\mu\nu}\,,
}
where $a,b$ indices are contracted with $\eta_{ab}$ and the vectors are written in polar coordinates.\footnote{When doing computations one has to be careful about how to compute the extrinsic curvature, as one needs to have a definition of $n^{(a)}$ away from the surface.}
Plugging in everything into the area functional \eqref{4derRT2} the equation of motion is obtained by varying. We get a complicated equation of motion, then we zoom in onto the near horizon region by taking  $z(r)=1-\ep\,s(r)^2$ and expanding for small $\ep$. To leading order in $\ep$, we get
\es{ZEoM}{
0=&\le(1+C_1(\Lam_1,\Lam_2)\ri)s''(r)-f_1\le({d-1\ov 2}+C_2(\Lam_1,\Lam_2)\ri)s(r)\\
&+ \Lam_2\, {s(r)\, s''(r)^2+2 s(r)\, s'(r) \, s^{(3)}(r)-2s'(r)^2\, s''(r)\ov s(r)^2}+O\le({1\ov r}\ri)\,,
}
where we have used the change of basis in the coupling constants \eqref{4derRT}, and there are many terms hidden in $O\le({1\ov r}\ri)$, including some that depend on $\Lam_\text{GB}$. These terms however will not play a role in the following, as we only want to determine $\tilde\mu$ from the large $r$ asymptotic behavior of the solution:
\es{Larger}{
s(r)\sim {e^{\tilde\mu r}\ov r^\#}\,.
}
From \eqref{ZEoM} it follows that we get the same equation for $\tilde\mu^2$ as we got for $\mu^2$, \eqref{muSq}, and we have to take the same root for $\tilde\mu^2$ that we took for $\mu^2$, as explained below \eqref{muSq}. Thus, using that $v_B={f_1\ov 2 \mu}$ and $\tilde v_B={f_1\ov 2 \tilde\mu}$, we conclude that $v_B=\tilde v_B$ in four derivative gravity to all orders in the higher curvature couplings. We note that the extrinsic curvature terms in \eqref{4derRT2} played an important role in the above computation.

\section{Bounds and the quasiparticle model}

We can also apply the bounds discussed in section \ref{boundsSec} to the quasiparticle model \cite{Calabrese:2005in,Calabrese:2007rg,Casini:2015zua,Cotler:2016acd} of integrable field theories. In these theories there is no chaos, hence no $v_B$ cone, and $v_{LC} = c$. For such systems, we do not expect the time dependence of the entropy to approximately saturate the bounds.

Let us first consider strip geometries of width $2R$ for $d>2$. For early times $t<R$:
\es{qpEarly}{
 S[A(t)]=v_E\, s_{\text{th}}\cdot\text{area}(A)\,t\,,
 }
 but for $t>R$ the result (computed in \cite{Casini:2015zua}) significantly deviates from the combined bound, in particular we find that $t_S=\infty$. This is in stark contrast with the holographic results, where \eqref{qpEarly} holds for all times until saturation, and $t_S=R/v_E$.\footnote{Another notable difference is that $v_E^\text{(free)}<v_E^\text{(holographic)}$, one of the main findings of  \cite{Casini:2015zua}.}
 
\begin{figure}[!h]
\begin{center}
\includegraphics[scale=.5]{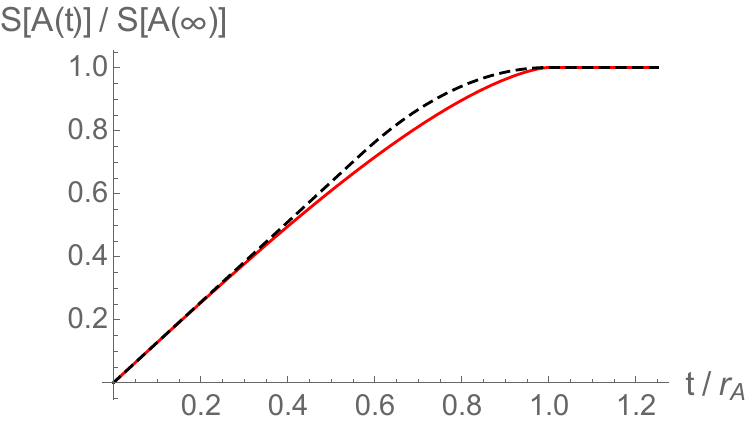}\hspace{0.3cm}
\includegraphics[scale=.5]{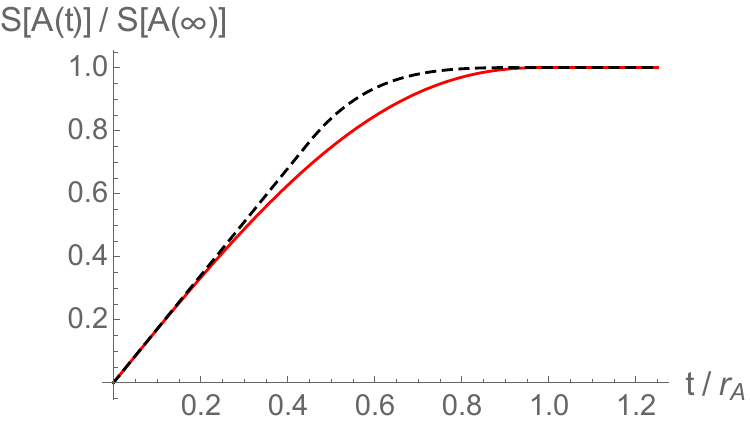}
\caption{We plot $S[A(t)]$ until saturation for a sphere in the quasiparticle model with EPR pattern of entanglement in $d=3,\,5$ respectively. Black/dashed is the upper bound \eqref{boundS} using the quasiparticle value of $v_E={2\ov\pi},\, {4\ov 3\pi}$ and with $\vv$ set to $c$, and red/solid is the quasiparticle result \cite{Casini:2015zua}.}\label{HcomparisonQP}
\end{center}
\end{figure}

However, somewhat coincidentally, when we consider spherical geometries the quasiparticle model comes closer to saturating the bounds. The expressions for $S[A(t)]$ in the quasiparticle model with EPR pattern of entanglement are given in \cite{Casini:2015zua}. In figure \ref{HcomparisonQP} we compare the results of the quasiparticle computation in $d=3,\,5$ to the combined bound. The curves are actually somewhat close. For example, the saturation time $t_S=R$ saturates the first bound in \eqref{satBound}.

We conclude that the quasiparticle model obeys bounds discussed in section \ref{boundsSec}. While the overall shape of the curves on figures \ref{Hcomparison} and \ref{HcomparisonQP} are similar, the finer details of entropy spread in chaotic systems is different from what one gets in the quasiparticle model. For generic shapes, we do not expect the quasiparticle model to approximately saturate the bounds from section \ref{boundsSec}.

\mciteSetMidEndSepPunct{}{\ifmciteBstWouldAddEndPunct.\else\fi}{\relax}
\bibliographystyle{utphys}
\bibliography{short.bib}{}

\end{document}